\documentclass[namedreferences]{solarphysics}
\usepackage[optionalrh]{spr-sola-addons} 
\usepackage{epsfig,color}          
\usepackage{url}             

\graphicspath{{figs/},{./}}

\newcommand{\pr}{{\mathrm{Pr}\,}}
\newcommand{\etasurf}{\eta_{\mathrm{surf}}}
\newcommand{\etaSCZ}{\eta_{\mathrm{SCZ}}}
\newcommand{\etaRZ}{\eta_{\mathrm{RZ}}}
\newcommand{\rBCZ}{r_{\mathrm{BCZ}}}
\newcommand{\rsurf}{r_{\mathrm{surf}}}
\newcommand{\OmegaSCZ}{\Omega_{\mathrm{SCZ}}}
\newcommand{\OmegaRZ}{\Omega_{\mathrm{RZ}}}
\newcommand{\OmegaEQ}{\Omega_{\mathrm{EQ}}}
\newcommand{\ov}{\overline}

\newcommand{\citet}{\inlinecite}

\def\NoBlackBoxes{\global\overfullrule 0pt}
\NoBlackBoxes

\begin{document}

\begin{opening}

\title{On the compatibility of a flux transport dynamo with a fast 
tachocline scenario}

\author{Bidya Binay \surname{Karak}$^{1}$\sep
        Kristof~\surname{Petrovay}$^{2}$      
       }
\runningauthor{B.B. Karak, K. Petrovay}
\runningtitle{Flux transport dynamo with a fast tachocline}

   \institute{$^{1}$~Department~of~Physics,~Indian~Institute~of~Science,~Bangalore~560012,~India
                     email: \url{bidya_karak@physics.iisc.ernet.in} \\ 
              $^{2}$~E\"otv\"os~University,~Department~of~Astronomy,~Budapest,~Hungary
                     email: \url{K.Petrovay@astro.elte.hu} \\
             }

\begin{abstract}
The compatibility of the fast tachocline scenario with a flux transport
dynamo model is explored. We employ a flux transport dynamo model 
coupled with simple feedback formulae relating the
thickness of the tachocline to the amplitude of the magnetic field or to
the Maxwell stress. The dynamo model is found to be robust against the
nonlinearity introduced by this simplified fast tachocline mechanism.
Solar-like butterfly diagrams are found to persist and, even without any
parameter fitting, the overall thickness of the tachocline is well
within the range admitted by helioseismic constraints. In the most
realistic case of a time and latitude dependent tachocline thickness 
linked to the value of the Maxwell stress, both the thickness and its
latitude dependence are in excellent agreement with seismic results. In
the nonparametric models, cycle related temporal variations in
tachocline thickness are somewhat larger than admitted by helioseismic
constraints; we find, however, that introducing a further parameter into
our feedback formula readily allows further fine tuning of the thickness
variations.
\end{abstract}
\end{opening}

\section{Introduction}
     \label{sect:intro} 

Flux transport dynamos are the most widely discussed scenario for the
origin of solar activity. In these models, the $\alpha$-effect
responsible for toroidal $\rightarrow$ poloidal flux conversion is
identified with the tilting of the axis of active regions relative to
the azimuthal direction. The poloidal field generated by this
$\alpha$-effect concentrated near the surface is then transported by the
meridional circulation to high latitudes and then down to the bottom of
the convective zone (Babcock--Leighton mechanism). The return branch of
a one-celled circulation flow pattern will then advect the field back
towards the equator while differential rotation amplifies it to a
toroidal field of increasing strength ($\Omega$-effect). When the
toroidal field strength reaches a critical limit, flux emergence driven
by the Parker instability will give rise to active regions on the
surface, with an axis tilted away from the azimuthal plane due to the
Coriolis force, thereby closing the cycle (see e.g. Charbonneau, 2010 
for a more detailed discussion). An attractive feature of this dynamo
model is that it not only models the regular features of the solar cycle
(Choudhuri, Sch\"ussler, and Dikpati, 1995; Dikpati and Charbonneau, 1999; Chatterjee, Nandy, and Choudhuri, 2004)
but also reproduces many irregular features of solar cycle (Charbonneau
and Dikpati, 2000; Choudhuri and Karak, 2009; Karak, 2010; Karak and 
Choudhuri, 2011; Choudhuri and Karak, 2012).

In order to avoid excessive buoyant loss of toroidal flux, the
$\Omega$-effect must take place in a stably stratified layer just below
the convective zone. Helioseismic inversions (Charbonneau et al., 1999; Basu and Antia, 2001)
indicate that this layer coincides with the solar tachocline, i.e.\ the 
thin transitional layer between the rigidly rotating interior and the 
differentially rotating convective zone. The narrow radial extent of 
this layer (scale depth of 5--10~Mm or less) implies a strongly 
anisotropic angular momentum transport mechanism, much more effective 
in the horizontal direction than in the radial direction. Spiegel and Zahn (1992) 
have shown that this is possible by taking strong anisotropic turbulent 
viscosity in the horizontal direction. However several authors 
(e.g., Gough and McIntyre, 1998; Dikpati and Gilman, 1999) have found caveats 
in this purely hydrodynamical model and improved by including additional 
physics. Other most plausible candidate for such anisotropic momentum transfer 
is the Maxwell stress in a predominantly horizontal magnetic field configuration.
One school of thought in this stream is that a weak fossil magnetic field in the
radiative zone may solve this problem (Rudiger and Kitchatinov, 1997; Gough and McIntyre, 1998; 
MacGregor and Charbonneau, 1999). Unfortunately the simulations (Garaud, 2002; 
Brun and Zahn, 2006; Strugarek, Brun, and Zahn, 2011) of this so-called slow 
tachocline model are not able to provide uniform rotation in the radiative 
zone even by including penetrating flows from the convection zone, such as 
plumes or meridional circulation to confine the fossil magnetic field 
(however Garaud and Garaud, 2008 succeeded in this problem). 
Other school of thought is that as the dynamo generated toroidal field resides in 
the tachocline, it is plausible to assume that this field is responsible for
the confinement of the tachocline. The feasibility of this so-called 
fast tachocline scenario has been demonstrated by 
Forg\'acs-Dajka and Petrovay (2001, 2002) and Forg\'acs-Dajka (2003).

In the fast tachocline scenario, the thickness of the tachocline depends
on the magnetic field. This thickness, on the other hand, is one input
parameter of flux transport dynamo models and thus it may be expected to
influence the amplitude and configuration of the magnetic field. This
coupling introduces a nonlinearity into the dynamo that may affect both
the dynamo and the structure in a number of ways. Indeed, it is not a
priori clear whether the fast tachocline scenario and flux transport
dynamos are compatible at all, i.e.\ whether they can give rise to a
finite amplitude oscillatory field with a tachocline of realistic
thickness and with cyclic variations within the observational bounds. 

The aim of the present paper is to investigate this issue. In a
completely self-consistent approach one should couple the angular
momentum transfer equation to the induction equation solved in dynamo
models. Instead of this more generic approach, as a first step, here we
parameterize the dependence of tachocline width $d_t$ on the poloidal
field amplitude $B_p$ with simple algebraic formulae that still retain
the essential physics of the problem. Details of our problem setup are
presented in Section~\ref{sect:setup}. Section~\ref{sect:results}
presents the results which are then confronted with observations in
Section 4. Section~\ref{sect:concl} concludes the paper.

\section{Problem setup} 
      \label{sect:setup}      

\subsection{Flux transport dynamo model}
An axisymmetric magnetic field can be represented in the form
\begin{equation}
{\bf B} = B (r, \theta) {\bf e}_{\phi} + \nabla \times [ A(r, \theta) {\bf e}_{\phi}],
\end{equation}
where $B (r, \theta)$ and $A(r, \theta)$ respectively correspond to the
toroidal and poloidal components. Then the evolution of magnetic fields
in  the flux transport dynamo model are governed by the following two
equations:

\begin{equation}
\frac{\partial A}{\partial t} + \frac{1}{s}({\bf v}.\nabla)(s A)
= \eta \left( \nabla^2 - \frac{1}{s^2} \right) A + S(r, \theta; B),
\end{equation}

\begin{equation}
\frac{\partial B}{\partial t}
+ \frac{1}{r} \left[ \frac{\partial}{\partial r}
(r v_r B) + \frac{\partial}{\partial \theta}(v_{\theta} B) \right]
= \eta \left( \nabla^2 - \frac{1}{s^2} \right) B 
+ s({\bf B}_p.{\bf \nabla})\Omega + \frac{1}{r}\frac{d\eta}{dr}\frac{\partial{(rB)}}{\partial{r}},~~~~~~~~~~~~~~~~~~~~~~~~~~~~~
\end{equation}

with $s = r \sin \theta$.

Here ${\bf v}$ is the meridional flow which has the following analytical form.
\begin{equation}
v_r(r,\theta)= \frac{v_0}{f} \left(\frac{R}{r}\right)^2 \left[ \frac{-1}{m+1}+\frac{c_1}{2m+1}\xi^m- \frac{c_2}{2m+p+1}\xi^{m+p} \right] \xi[2\cos^2\theta-\sin^2\theta]\label{mc1}
\end{equation}

\begin{equation}
v_\theta(r,\theta) = \frac{v_0}{f} \left(\frac{R}{r}\right)^3 [-1+c_1\xi^m-c_2\xi^{m+p}] \sin\theta\cos\theta,\label{mc2}
\end{equation}
with
 $\xi(r)=\frac{R}{r}-1$, $c_1=\frac{(2m+1)(m+p)}{(m+1)p}\xi^{-m}_p$, $c_2=\frac{(2m+p+1)m}{(m+1)p}\xi^{-(m+p)}_p$ and $\xi_p=\frac{R}{r_p}-1$.
Here $m=0.5$, $p = 0.25$, $v_0 = 24$~m~s$^{-1}$ and $f$ is the normalization factor which
determines the maximum value of the latitudinal component of the meridional circulation $v_\theta$.

$\eta$ is the turbulent magnetic diffusivity which has the following form:
\begin{equation}
\eta(r) = \etaRZ + \frac{\etaSCZ}{2}\left[1 + \mathrm{erf} \left(2\frac{r - \rBCZ}
{d_t}\right) \right]+\frac{\etasurf}{2}\left[1 + \mathrm{erf} \left(\frac{r - \rsurf}
{d_2}\right) \right]
\label{eq:etap}
\end{equation}
with $\rBCZ=0.7R$, $d_t=0.03R$, $d_2=0.05R$, $\rsurf = 0.95R$, 
$\etaRZ = 5 \times 10^8$ cm$^2$~s$^{-1}$, $\etaSCZ = 5 \times 10^{10}$ cm$^2$~s$^{-1}$ and 
$\etasurf = 2\times10^{12}$ cm$^2$ s$^{-1}$.

$S(r, \theta; B)$ is the coefficient which describes the generation
of poloidal field at the solar surface from the decay of tilted bipolar
sunspots (Babcock-Leighton process). It has the following form
\begin{equation}
 S(r, \theta; B) = \frac{\alpha(r,\theta)}{1+(B(r_t,\theta)/B_0)^2} B(r_t,\theta),
\label{alpha1}
\end{equation}
where
$\alpha(r,\theta)=\frac{\alpha_0}{4}\left[1+\mathrm{erf}\left(\frac{r-r_4}{d_4}\right)\right]\left[1-\mathrm{erf}\left(\frac{r-r_5}{d_5}\right)\right]\sin\theta\cos\theta\left[\frac{1}{1+e^{-\gamma(\theta-\pi/4)}}\right]$
with $r_4=0.95R$, $r_5=R$, $d_4=0.05R$, $d_5=0.01R$,
$\gamma=30$, $\alpha_0 = 1.6$ cm~s$^{-1}$ and $B_0 = 4 \times 10^4~G$.

$\Omega$ is the solar rotation which has the following form
\begin{equation}
\Omega(r,\theta) = \OmegaRZ + \frac{1}{2}\left[1 + erf \left(2\frac
{r - r_t}{d_t}\right) \right]\left[\OmegaSCZ(\theta) - \OmegaRZ\right],
\label{eq:omegaprofile}
\end{equation}
where $r_t$ (the position of the tachocline) $= 0.7R$, $d_t$
(the thickness of the tachocline) $= 0.03R$,
$\OmegaRZ/2\pi = 432.8$ nHz, $\OmegaSCZ(\theta) = \OmegaEQ 
+ \alpha_2 cos^2(\theta) + \alpha_4 cos^4(\theta)$, with $\OmegaEQ/2\pi 
= 460.7$ nHz, $\alpha_2/2\pi = -62.69$ nHz and $\alpha_4/2\pi = -67.13$ nHz.

With the appropriate boundary conditions we solve Equations (2) and (3) in a full 
sphere of the meridional plane with $0.55R < r < R$, $0 <\theta <\pi$
to study the evolution of magnetic fields (see Chatterjee, Nandy, and Choudhuri, 2004 
for details).

\begin{figure}
\begin{center}
\epsfig{width=\textwidth,figure=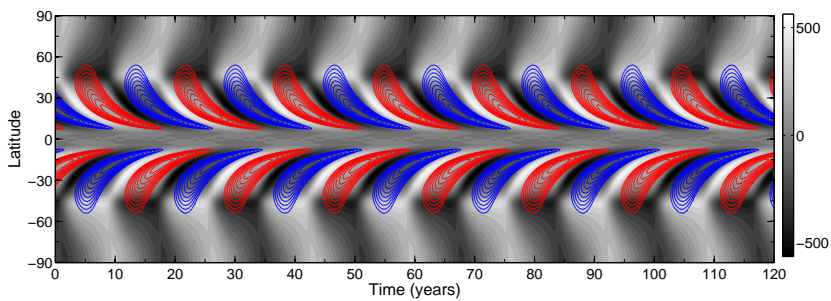}
\caption{Reference model with fixed tachocline thickness $d_t$. Contours
show the butterfly diagram of the toroidal field in the tachocline, the
field strength being $0.91\times 10^5 G$ at the innermost contours and
half that value at the outermost contours. Blue contours correspond to
positive toroidal field whereas red contours correspond to negative
toroidal field. The greyscale background shows the weak diffuse radial
field on the solar surface.}
\label{reference}
\end{center}
\end{figure}

Once the transient state is over the butterfly diagram looks like the
one shown in Figure \ref{reference}. In this figure the contours show
the toroidal field in the tachocline and the shaded background shows the
strength of the radial  field on the solar surface.

In the next section, we shall discuss the dependence of the tachocline 
thickness on the magnetic field based on the fast tachocline model and
then  we shall study its effect on this flux transport dynamo model.

\subsection{Tachocline depth parametrization}  

Tachocline thickness is introduced in various ways by various authors,
so care must be taken when comparing results from different papers. For
a comparison it is useful to define the {\it relative residual angular
velocity} $\tilde\Omega$ as
\begin{equation} 
\tilde\Omega (r,\theta) = \frac{\Omega(r,\theta)-\OmegaRZ}{\OmegaSCZ(\theta) 
-\OmegaRZ}.
\end{equation} 
Clearly, $\tilde\Omega$ takes the value $0$ in the radiative zone, 1 at the
top of the tachocline. The radius $r_t$ of the tachocline is defined by
$\tilde\Omega(r_t,\theta)=0.5$.

If $\tilde\Omega$ varies by a factor $F$ in the radius range
$[r_t-x/2,r_t+x/2]$, then clearly $x/H=\ln F$ where $H$ is the scale
height of the tachocline. From Equation (\ref{eq:omegaprofile}) it then
follows that our tachocline thickness parameter $d_t=1.65 H$. Similarly,
for the tachocline thickness parameter $w$ used by Antia and Basu (2011)
in their helioseismic study of tachocline properties we find $w=0.39H$,
i.e. $d_t\simeq 4.3w$. This relation must be taken into account when
comparing our results to observational constraints.

An approximate analytic formula relating the mean tachocline scale
height $H$ to the amplitude of an imposed poloidal field in a periodic shear
layer was derived by Forg\'acs-Dajka and Petrovay (2001).  It reads 
\begin{equation} 
  {V_p^2}= \frac{\pr\eta r_t^2\omega}{H^2}  
  \frac{(1+\eta/\omega H^2)(1+\pr\eta/\omega H^2)}{1+2\pr\eta/\omega H^2},
  \label{eq:dVprel} 
\end{equation} 
where $\omega=2\pi/22$ years is the dynamo frequency, $\pr$ is the
Prandtl number, $\eta$ is the turbulent magnetic diffusivity, 
$r_t$ is the radius of the tachocline and $V_p$ is the
amplitude of the oscillatory poloidal magnetic field in Alfv\'enic
units. In a turbulent medium the effective Prandtl number is expected to
be order of unity, while its exact value is unknown. Therefore, in what
follows we shall simply substitute $d_t$ for $H$ in Equation
(\ref{eq:dVprel}), ignoring the factor 1.65 between these scales as an
appropriate choice of $\pr$ can always offset this factor anyway.

\begin{figure}
\begin{center}
\epsfig{width=0.6\textwidth,figure=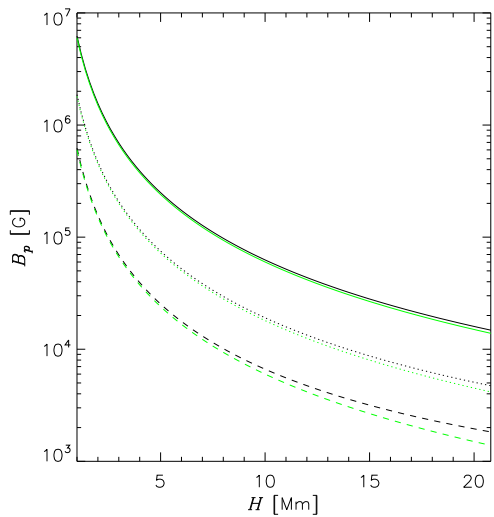}
\caption{Relation between tachocline width $d_t$ and poloidal field
amplitude $B_p$ for three different values of magnetic diffusivity,
$\eta=10^{12}$ (solid), $\eta=3\times 10^{11}$ (dotted), $\eta=10^{11}$
(dashed). Green curves show the function $B_p= 6\times 10^{10} \eta/d_t^2$.}
\label{fig:dBprel}
\end{center}
\end{figure}

Figure \ref{fig:dBprel} presents the variation of $B_p$  (= $V_p (4 \pi
\rho)^{1/2}$ with $\rho = 0.1$~g/cm$^3$) based on the relation
(\ref{eq:dVprel}) for three different values of the diffusivity. Green
curves are simpler analytical fits to each curve with a function of the form
${B_p}= {C \eta}/{d_t^2}$ or 
\begin{equation}
  d_t^2 = \frac{C \eta} {B_p},
  \label{eq:fits}
\end{equation}
where $C = 6\times 10^{10}$\,G$\cdot$s, and $B_p = \sqrt{(B_r^2 +
B_\theta^2)}$.

In a first attempt to relate tachocline thickness to the dynamo field by
a simple power law formula, in Section \ref{sect:results1} we shall employ
formula (\ref{eq:fits}) as a guidance in our choice of the exponent. It
should be emphasized that, as Equation (\ref{eq:fits}) refers to cycle averaged
quantities, it should not be viewed as rigorous derivation of our coupling
formulae (\ref{eq:fitsnew}) and (\ref{eq:fitsnew2})). Rather, it provides a
distant analogy guiding us in our choice of a physically motivated exponent
value. In particular, in the fast tachocline scenario the tachocline is
confined by the Maxwell stress ${B_p B}$ which also involves the toroidal field
strength $B$. The reason why $B$ does not explicitly appear in Equation
(\ref{eq:fits}) is that the toroidal field is generated by the windup of the
poloidal field, so its {\it overall} amplitude is ultimately determined by
$B_p$. However, this is clearly only true for the overall amplitudes and not for
the actual values of $B_p$ and $B$ at any given point and instant: so the 
simple approximate relationship (\ref{eq:fits}) is, strictly speaking, only
valid for the (latitudinal and temporal) average field amplitude and average
tachocline thickness and its application to relate $d_t$ and $B_p$ at any given
point in space and time is not well justified.

For this reason, in Section \ref{sect:results2} we shall try to further
improve on the model by relating $d_t$ directly to the Maxwell stress.
We recall from Forg\'acs-Dajka and Petrovay (2001) that the simple model giving rise to Equation (\ref{eq:fits})
consists of a semi-infinite layer bounded from above at depth $z=0$
(``bottom of the convective zone'') where a periodic shear flow is
imposed in the $y$ direction:
\begin{equation} 
  v_{y0}= v_0 \cos(kx). 
\end{equation} 
An oscillatory horizontal ``poloidal'' field is prescribed in the $x$
direction throughout the volume as
\begin{equation} 
  B_x=B_p \cos(\omega t). 
\end{equation} 
Introducing $v=v_y$ and using Alfv\'en speed units for the magnetic field
\begin{equation} 
  V_p=B_p(4\pi\rho)^{-1/2} \qquad b= B_y(4\pi\rho)^{-1/2}, 
\end{equation} 
the azimuthal component of the equation of motion reads 
\begin{equation} 
  \partial_t v=V_p \cos(\omega t)\partial_x b-\nu\nabla^2 v , \label{eq:veq} 
\end{equation} 
where $\nu$ denotes viscosity. Solutions may be sought in the form
\begin{equation} 
   v=\ov v(x, z) +v'(x,z) f(\omega t),  \qquad 
   b=b'(x,z) f(\omega t+\phi),         
\end{equation} 
where $f$ is a $2\pi$-periodic function of zero mean and of amplitude
${\cal O}(1)$. ($\ov a$ denotes time average of $a$, while $a'\equiv
a-\ov a$.) The (temporal) average of Equation (\ref{eq:veq}) then reads
\begin{equation} 
  0=V_p \ov{\cos(\omega t)f(\omega t+\phi)} \partial_x b' -\nu\nabla^2 \ov v. 
   \label{eq:vmean}  
\end{equation} 

For an order of magnitude estimate of the terms of this equation, we
assume $\ov{\cos(\omega t)f(\omega t+\phi)}\in{\cal O}(1)$ (i.e.\ no
``conspiracy'' between the phases, a rather natural assumption). As
$H\ll R$ we may approximate $\nabla^2\sim H^{-2}$. Estimating the other
derivatives as $\partial_t\sim\omega$ and $\partial_x\sim R^{-1}$,
(\ref{eq:vmean}) yields
\begin{equation}  
  V_p b'/R \sim \nu \ov v/H^2 ,  \label{eq:vmeanest}  
\end{equation} 
i.e.,
\begin{equation} 
  H^2 \simeq \frac{\nu R\ov v}{V_p b'} 
\end{equation} 
or, switching back to conventional magnetic field units and from $H$ to
$d_t$: \begin{equation} d_t^2 = \frac{C'\eta}{B_p B} . 
  \label{eq:fits2}
\end{equation}
Plugging in the correct dimensional values we find 
$C'\simeq 2\times 10^{15}\,$G$^2$$\cdot$s 
---this is the value we use in Section \ref{sect:results2} below where,
again, we shall employ formula (\ref{eq:fits2}) as a guidance in our
choice of the exponent of our simple feedback formula. 
While the link between tachocline thickness and Maxwell stress is more direct
than between $d_t$ and $B_p$ alone, it should still be kept in mind that
equation (\ref{eq:fits2}) refers to cycle averaged quantites, so it should not
be viewed as a rigorous derivation of our coupling formulae (\ref{eq:fits2new})
and (\ref{eq:fits2new2}). Rather, it provides an analogy guiding us in our
choice of a physically motivated exponent value.

\section{Results} 
      \label{sect:results}      

\subsection{Tachocline thickness linked to poloidal field amplitude} 
      \label{sect:results1}      

In the fast tachocline scenario the thickness of the tachocline may vary
as a function of both time and latitude. For a fully consistent treatment
of this variation the equation of motion should be coupled to the dynamo
equations. However, in the present preliminary exploration of the
problem we are only interested in the general stability properties and
robustness of flux transport dynamo against the kind of nonlinearity
introduced by the fast tachocline mechanism. Therefore, we simplify the
problem by an appropriate parametrization of the dependence of $d_t$ on
the magnetic field. For this purpose any arbitrary numerical relation
between tachocline thickness and magnetic field may do as long as it
results in $d_t$ values of the right order of magnitude and it has the
expected property of resulting in a thinner tachocline for stronger
magnetic fields. 

The numerical relation we first consider here is Equation
(\ref{eq:fits}). As we already remarked in Section 2.2, by its derivation,
the formula (\ref{eq:fits}) given above only relates the cycle and
latitude averaged mean value of the tachocline thickness to the
amplitude of the variation of the poloidal field, and our use of it to
link local and momentary values of these variables is arbitrary.
Nevertheless, by its construction, Equation (\ref{eq:fits}) does have
the required properties of resulting in $d_t$ values of the right order
of magnitude and in a thinner tachocline for stronger magnetic fields,
so it will suffice for the purpose of a first exploration.

For a first study we shall not consider any latitudinal variation of the
tachocline. Substituting the appropriate turbulent value for $\eta$ in the
tachocline we then have
\begin{equation} 
  {d_t^2} = \frac{{C \eta_t}} {\bar B_p(t)}
  \label{eq:fitsnew} 
\end{equation}
where
\begin{equation} 
  \eta_t = \frac 1{2d_0} \int_{r_t-d_0}^{r_t+d_0} \eta(r)\,dr 
\end{equation}
with $d_0=0.015R$. 
\begin{equation} 
  \bar B_p (t) = \frac 2{\pi} \int_0^{\pi/2} \bar B_p (\theta,t)
  \,d\theta
\end{equation}
is the latitude averaged poloidal field amplitude while  $\bar B_p
(\theta, t)$ is the local radial mean value of the poloidal field
calculated as
\begin{equation} 
  \bar B_p (\theta, t) = \frac 1{2d_0} \int_{r_t-d_0}^{r_t+d_0} \bar B_p
  (r,\theta, t)\,dr .
\end{equation}
(Recall that $\bar B_p = \sqrt{(B_r^2 + B_\theta^2)}$, i.e. the modulus of the
poloidal field. This ensures that $\bar B_p$, as given in the above equations,
remains positive at all times.)

\begin{figure}
\begin{center}
\epsfig{width=\textwidth,figure=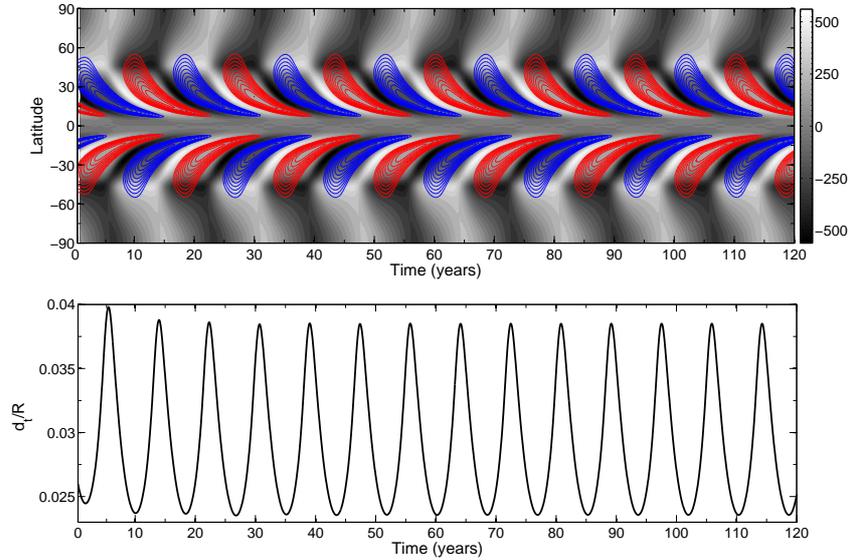}
\caption{Upper panel: same as Figure \ref{reference} but for variable
tachocline thickness $d_t$ as given by eq. (\ref{eq:fitsnew}).  Lower
panel: variation of tachocline thickness $d_t$ with time.}
\label{tactt}
\end{center}
\end{figure}

\begin{figure}
\begin{center}
\epsfig{width=\textwidth,figure=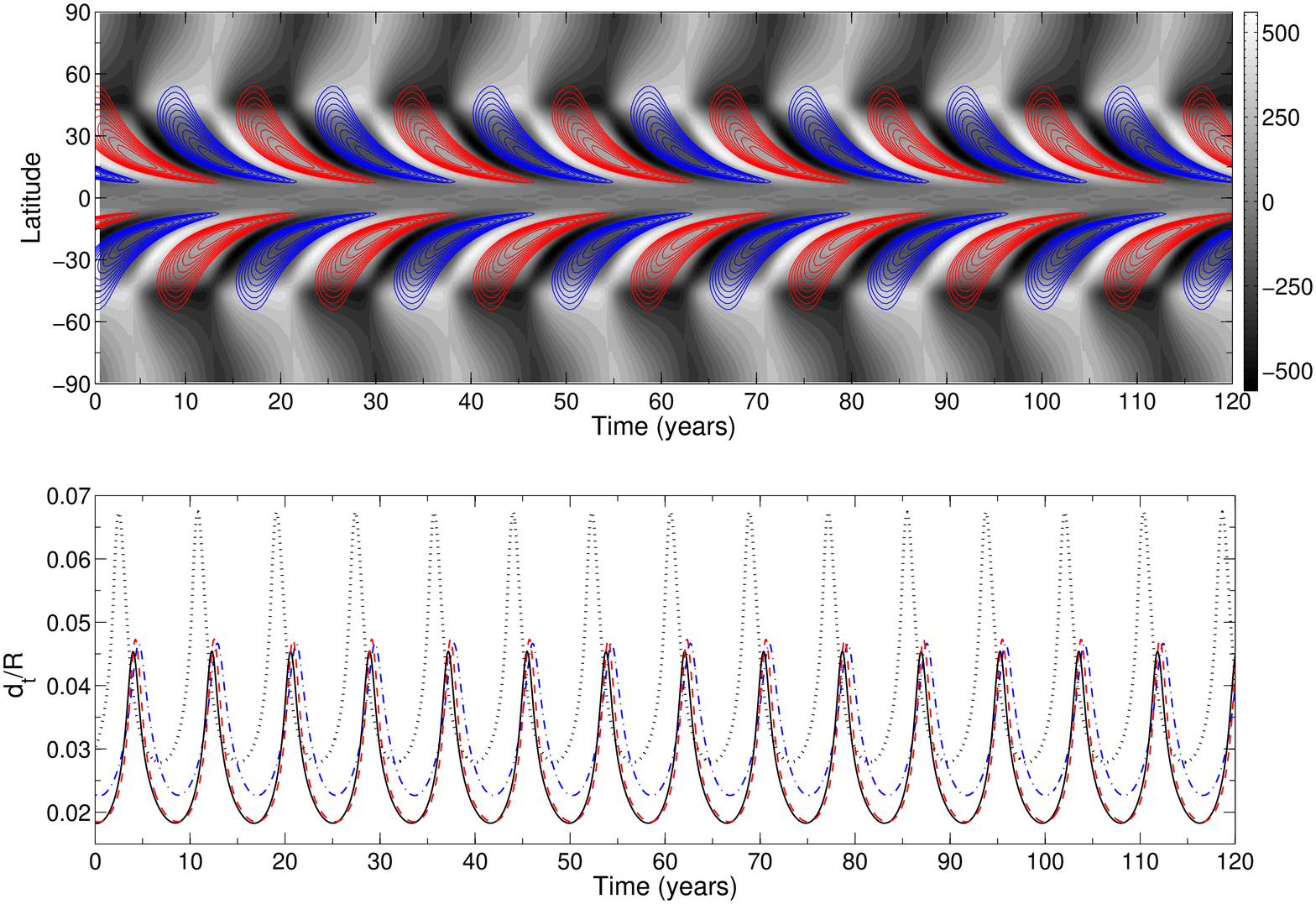}
\caption{Upper panel: same as Figure \ref{reference} but for variable
tachocline thickness $d_t$ as given by eq. (\ref{eq:fitsnew2}). Lower
panel: variation of tachocline thickness $d_t$ with time at different
latitudes. Dash-dotted, solid, dashed and dotted lines are the values at
$75^0$, $60^0$,  $45^0$ and $15^0$ latitudes, respectively.}
\label{tacttheta}
\end{center}
\end{figure}

Now, starting from the relaxed state of our (fixed $d_t$) model
discussed in Section~2.1, we allow $d_t$ to vary at every time step of our
simulation, calculating $\bar B_p(t)$ as above to get the value
of ${d_t}$ from Equation (\ref{eq:fitsnew}).  The calculation is run
again for several solar cycles until it relaxes to a nearly steady
cyclic behavior. In Figure \ref{tactt}, we show a few solar cycles
long clip just after this modification. It is noteworthy that this kind
of tachocline structure calculated from the fast tachocline model does
produce a periodic solar-like solution. In the upper panel, the butterfly
diagram of toroidal and radial field components is shown. The appearance
of this plot is quite similar to the one found in Figure 
\ref{reference} in Section~2.1. Therefore we may conclude that the fast 
tachocline model and the flux transport dynamo model are compatible and
a nonlinear coupling between tachocline thickness and field amplitude
does not lead to a breakdown of the solar-like solution.

The lower panel shows the variation of ${d_t}$ with
solar  cycle. The amplitude of this variation is about $\pm 25\,$\%.

Next, we try to explore the latitudinal structure of the tachocline by
admitting $d_t$ to depend on $\theta$, i.e. instead of Equation
(\ref{eq:fitsnew}) we use
\begin{equation} 
  {d_t^2} = \frac{{C \eta_t}} {\bar B_p(\theta, t)} .
  \label{eq:fitsnew2} 
\end{equation}

Figure \ref{tacttheta} presents the result for this calculation. Upper
panel shows the butterfly diagram which still resembles most
characteristics of the observed butterfly diagram.

\subsection{Tachocline thickness linked to the Maxwell stress} 
      \label{sect:results2}      

In our next parametric study we link the tachocline thickness $d_t$ to
the Maxwell stress by formula (\ref{eq:fits2}). Again, for our first
study we disregard the latitude dependence in $d_t$:
\begin{equation} d_t^2 = \frac{C'\eta_t}{\bar B_p(t) \bar B(t)}  
  \label{eq:fits2new}
\end{equation}
The mean values of $B$ are here defined in a manner analogous to those
of $B_p$:
\begin{equation} 
  \bar B (t) = \frac 2{\pi} \int_0^{\pi/2} \bar B (\theta,t)
  \,d\theta
\end{equation}
is the latitude averaged toroidal field amplitude while  $\bar B
(\theta, t)$ is the local radial mean value of the toroidal field
calculated as
\begin{equation} 
  \bar B (\theta, t) = \frac 1{2d_0} \int_{r_t-d_0}^{r_t+d_0} |B
  (r,\theta, t)|\,dr .
\end{equation}
(Again, using the modulus of the toroidal field in the integrand ensures
that $\bar B$, as given in the above equations, remains positive at all
times.)

\begin{figure}
\begin{center}
\epsfig{width=\textwidth,figure=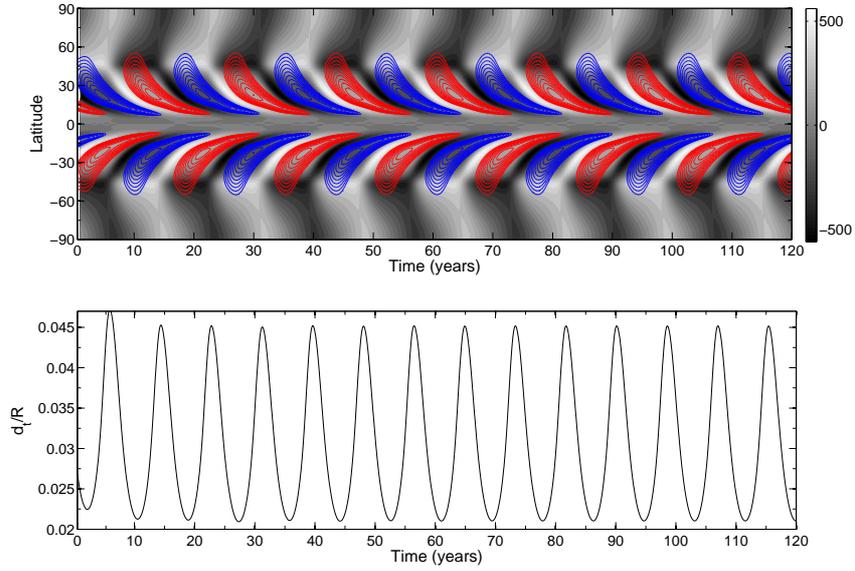}
\caption{Same as Figure \ref{tactt} but here the tachocline thickness $d_t$
is determined by eq. (\ref{eq:fits2new}).}
\label{maxt}
\end{center}
\end{figure}

The result is shown in Figure \ref{maxt}. 

\begin{figure}
\begin{center}
\epsfig{width=\textwidth,figure=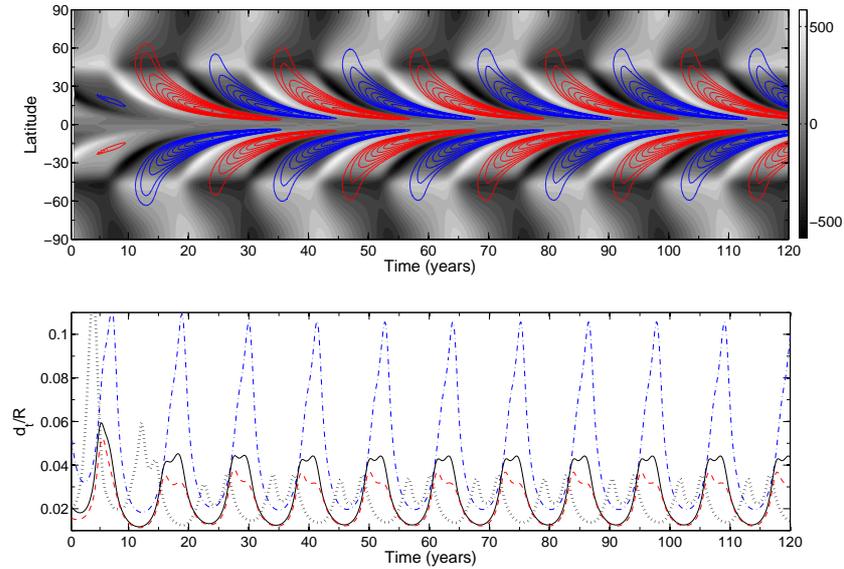}
\caption{Same as Figure \ref{tacttheta} but here the tachocline thickness $d_t$ 
is determined by eq. (\ref{eq:fits2new2}) (most realistic model). In the lower panel, 
the dash-dotted, solid, dashed and dotted lines are the values of $d_t$ 
at $75^0$, $60^0$,  $45^0$ and $15^0$ latitudes, respectively.}
\label{maxttheta}
\end{center}
\end{figure}

Finally, to explore the latitude dependence of tachocline thickness we set
\begin{equation} d_t^2 = \frac{C'\eta_t}{\bar B_p(\theta, t) 
  \bar B(\theta, t)}.  
  \label{eq:fits2new2}
\end{equation}
This yields what we may regard our most realistic model. Figure
\ref{maxttheta} shows the result.

\begin{figure}
\begin{center}
\epsfig{width=\textwidth,figure=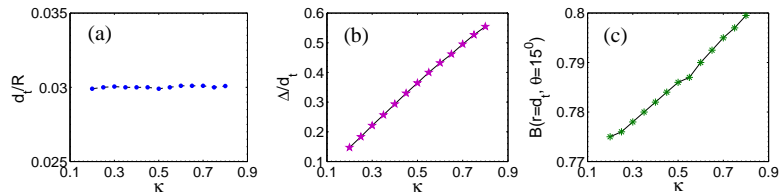}
\caption{Variation of tachocline thickness $d_t$ (a); of the amplitude
variation of the tachocline thickness $\Delta/d_t$ (b); and of toroidal
field strength at $15^\circ$ latitude (c) as functions of $\kappa$ (the
exponent in Equation \ref{eq:fitskappa}.} 
\label{exponent}
\end{center}
\end{figure}

\section{Discussion}
      \label{sect:discussion}      

The properties of the variable tachocline resulting from our models may
be compared to empirical (helioseismic) constraints. An in-depth study
of tachocline thickness as a function of latitude and time was recently
performed by  Antia and Basu (2011) (also see Basu and Antia, 2001).
Converting their tachocline thickness parameter $w$ to our parameter
$d_t$ as discussed in Section 2.2 above, their finding was that $d_t/R$
varies from approximately $0.02$ to $0.1$ with increasing heliographic
latitude. Temporal variations of up to $\pm 50\,$\% are observed;
however these are dominated by variation on the time scale of a few
years. (Variations on such time scales may be produced by shear
instabilities, cf. Miesch, 2007) On the longer time scale of 11 years no
systematic variations were detected; from visual inspection of the
relevant figures, this sets an upper limit of $\sim\pm 20\,$\% on cycle
related changes in tachocline thickness.

The mean value of tachocline thickness in Figures
\ref{tactt}--\ref{maxttheta} lies around  $0.03R$. In the most realistic
case, Figure \ref{maxttheta}, the thickness varies from $0.02$ to $0.1$
as we move from low to high latitudes, in perfect agreement with the
observations. This demonstrates that the simple analytical feedback
formula (\ref{eq:fits2new2}) not only gives a solar like dynamo solution
in the flux transport dynamo model but also reproduces the realistic
value {\it and} the observed latitude dependence of the tachocline
thickness. 

The cycle variation of tachocline thickness found in our models is quite
significant, up to $\pm 30\,$\%. This is somewhat higher than what the
observational constraints suggest.

In order to explore how sensitive our quantitative results are to
details of the feedback formula, we generalize Equation
(\ref{eq:fits2new2}) as
\begin{equation} d_t = \frac{(C'\eta_t)^{1/2}}{[\bar B_p(\theta, t) 
  \bar B(\theta, t)]^{\kappa}}.  
  \label{eq:fitskappa}
\end{equation}
Clearly, the case $\kappa=0.5$ returns Equation (\ref{eq:fits2}). While
using other values of $\kappa$ has no clear physical justification, it
offers a way to explore the sensitivity of our quantitative results to
details of the feedback formula chosen.

We repeat the same calculation at different values of $\kappa$ ranging
from 2 to 8. Figure \ref{exponent} presents the results. Panel (a)
shows the variation of ${d_t}$ as a function of $\kappa$. Note that  as we
vary $\kappa$, we also vary $C'$ in order to fix the mean value of
${d_t}$ at around $0.03R$ which is fairly close to the observed value.
(As $C'$ involves the unknown quantity $\pr$, its value is not well
constrained anyway.) Figure \ref{exponent}(b) shows the amplitude
$\Delta$  of the departure of $d_t$ from its mean value as a function of
$\kappa$. A smooth increase with $\kappa$ is found, so  lower $\kappa$
values result in rather more subdued thickness variations that may be
compatible with more stringent empirical limits. 
Following Choudhuri (2003) we should keep in mind that as we are using mean field dynamo 
equations, the Maxwell's stress $B_pB$ is consist of the mean values of $B_p$ and $B$ 
and thereby these are 
non-zero only inside the flux tubes. Now if $(B_p)$ft and $(B)$ft are 
the values of these quantities inside flux tubes and $f$ is the filling factor, 
then we have $B_p = f (B_p)$ft and $B = f (B)$ft (we assumed the same 
filling factor for both components for the sake of simplicity). 
Therefore it is easy to show that the mean stress appeared in all the calculations 
of Section~3.2 should be replaced by the 
$f (B_p)$ft $(B)$ft = $B_p B / f $
However this factor $f$ can be absorbed in the $C'$  because $C'$ is not a strict constant.

\section{Conclusion}
      \label{sect:concl} 

We have explored the compatibility of the fast tachocline model with the
flux  transport dynamo model. For this purpose we employed two simple
feedback formula, Equations (\ref{eq:fitsnew2}) and
(\ref{eq:fits2new2}), relating the thickness of the tachocline to the
poloidal magnetic field strength. While the second of these formulae is
physically more sound, both reflect the expected form of the
relationship between cycle averaged tachocline thickness and the
amplitude of the magnetic field, and they do have the required
properties of resulting in $d_t$ values of the right order of magnitude
and in a thinner tachocline for stronger magnetic fields.  Both cases
were studied with and without latitude dependence in $d_t$.

The flux transport dynamo model we considered proved to be robust
against the nonlinearity introduced by this simplified fast tachocline
mechanism. Solar-like butterfly diagrams were found to persist and the
overall thickness of the tachocline is well within the range admitted by
helioseismic constraints even without any parameter fitting. In the most
realistic case of a time and latitude dependent tachocline thickness
linked to the value of the Maxwell stress, $d_t$ and its latitude
dependence are in excellent agreement with seismic results. 

One characteristic of fast dynamo mechanism is a marked cycle dependence
in tachocline properties. Indeed, without parameter fitting, in all of
our models the cycle related changes in tachocline thickness are
somewhat larger than the maximal such variation admitted by helioseismic
constraints. On the other hand, an exploration of the parameter space
indicates that in a slightly generalized form of our feedback formula it
is possible to find parameter combinations where the cycle variation
remains within the observational bounds while in other respects the good
agreement of our model with observations is preserved.

While there is no straightforward physical justification for such a
generalized feedback formula, our parametric study shows that results
like the compatibility of the fast dynamo and flux transport dynamo
mechanism, the overall thickness of the tachocline or the amplitude of
the toroidal field are more robust than details like the time and
latitude dependence. Physical effects not considered in our model may,
then, potentially be held responsible for the deviation of the feedback
formula exponent from the value of 0.5 derived from a simplified
analytic model.

Such further effects may include tachocline instabilities, a change in
the geometric setup of our dynamo model or a combination of the two. One
clear limitation of the model is that differential rotation is quite
strong at high latitudes, so one may expect that this is the region
where strong toroidal field is produced.  However, sunspots appear at
low latitudes only. In the Surya family of models on which our model is
based, this problem is bypassed by considering a deeply penetrating
meridional flow (Nandy and Choudhuri, 2002). While this may be
unphysical, it has the desired effect of allowing the toroidal field
generated at higher latitudes to be stored and amplified further until
it reaches the lower latitudes where sunspot eruptions happen. A
promising, and more physically consistent way out of this conundrum was
suggested by Parfrey and Menou (2007) who showed that at latitudes
higher than 37$^\circ$ the magnetorotational instability (MRI) is
present in the tachocline, presumably resulting in stronger turbulence,
a thicker tachocline and a less organized field structure. All this just
serves to illustrate that, independently of the choice of our feedback
formula, there may be more physical effects to consider before we can
hope to simultaneously reproduce the correct temporal and latitudinal
structure of the tachocline.

A further obvious limitation is that only one particular dynamo model
was considered. It is well known that properties of flux transport
dynamos are quite sensitive to variations in the input parameters and
assumptions (e.g. Dikpati et al., 2002; Guerrero and  de Gouveia Dal Pino,
2007; Jiang, Chatterjee, and Choudhuri, 2007; Yeates, Nandy, and Mackay,
2008; Karak and Choudhuri, 2012; Karak and Nandy, 2012). It is therefore  desirable to extend
such studies to other dynamo configurations.

It is nevertheless clear that the use of simple feedback formulae of any
form via a simplified approach to the problem can only be considered as a
first exploration. In any case, the completely consistent
treatment of the problem should involve the full solution of the
equation of motion, coupled with the dynamo equations.

\begin{acks}
We thank Prof. Arnab Rai Choudhuri for raising valuable points which certainly 
improved this paper. For our dynamo calculations we have used Surya code developed 
in Choudhuri's group at Bangalore. Support by the Hungarian Science Research Fund (OTKA grants no.\ K83133
and 81421) and by the European Union with the co-financing of the
European Social Fund (grant no.\ T\'AMOP-4.2.1/B- 09/1/KMR-2010-0003) is
gratefully acknowledged.
BBK thanks CSIR, India for financial support and also to ELTE for comfortable
hospitality  during the beginning of this work.
\end{acks}


\begin{thebibliography}{}
\bibitem[Antia and Basu 2011]{ab11}Antia, H.M., Basu, S.: 2011, {\it Astrophys. J. Lett.} {\bf 735,} L45.
\bibitem[Brun and Zahn 2006]{brun06}Brun, A.S., Zahn, J.-P.: 2006, {\it Astron. Astrophys.} {\bf 457,} 665.
\bibitem[Basu and Antia 2001]{ba01}Basu, S., Antia, H.M.: 2001, {\it Mon. Not. Roy. Astron. Soc.} {\bf 324,} 498.
\bibitem[Charbonneau 2010]{c10}Charbonneau, P.: 2010, {\it Living Rev. Solar Phys.} {\bf 7,} 3.
\bibitem[Charbonneau and Dikpati 2000]{cd00}Charbonneau, P., Dikpati, M.: 2000, {\it Astrophys. J.} {\bf 543,} 1027.
\bibitem[Charbonneau et al. 1999]{charbonneau99}Charbonneau, P. Christensen-Dalsgaard, J., Henning, R., Larsen, R.M., Schou, J., Thompson, M.J. et al.: 1999, {\it Astrophys. J.} {\bf 527,} 445.
\bibitem[Chatterjee Nandy, and Choudhuri 2004]{cnc04}Chatterjee, P., Nandy, D., Choudhuri, A.R.: 2004, {\it Astron. Astrophys.} {\bf 427,} 1019.
\bibitem[Choudhuri 2003]{c03}Choudhuri, A.R.: 2003, {\it Solar Phys.} {\bf 215,} 31.
\bibitem[Choudhuri and Karak 2009]{ck09}Choudhuri, A.R., Karak, B.B.: 2009, {\it Res. Astron. Astrophys.} {\bf 9,} 953.
\bibitem[Choudhuri and Karak 2012]{ck12}Choudhuri, A.R., Karak, B.B.: 2012, {\it Phys. Rev. Lett.} submitted (arXiv:1208.3947).
\bibitem[Choudhuri, Sch\"ussler, and Dikpati 1995]{csd95}Choudhuri, A.R., Sch\"ussler, M. Dikpati, M.: 1995, {\it Astron. Astrophys. Lett.} {\bf 303,} L32.
\bibitem[Dikpati and Charbonneau 1999]{dc99}Dikpati, M., Charbonneau, P.: 1999, {\it Astrophys. J.} {\bf 518,} 508.
\bibitem[Dikpati and Gilman 1999]{dg99}Dikpati, M., Gilman, P.A.: 1999, {\it Astrophys. J.} {\bf 512,} 417.
\bibitem[Dikpati et al. 2002]{dik02} Dikpati, M., Corbard, T., Thompson, M.J., Gilman, P.A.: 2002, {\it Astrophys. J. Lett.} {\bf 575,} L41.
\bibitem[Forg\'acs-Dajka and Petrovay 2001]{petrovay01}Forg\'acs-Dajka, E., Petrovay, K.: 2001, {\it Solar Phys.} {\bf 203,} 195.
\bibitem[Forg\'acs-Dajka and Petrovay 2002]{petrovay02}Forg\'acs-Dajka, E., Petrovay, K.: 2002, {\it Astron. Astrophys.} {\bf 389,} 629.
\bibitem[Forg\'acs-Dajka 2003]{dajka03}Forg\'acs-Dajka, E.: 2003, {\it Astron. Astrophys.} {\bf 413,} 1143.
\bibitem[Garaud 2002]{garaud02}Garaud, P.: 2002, {\it Mon. Not. Roy. Astron. Soc.} {\bf 329,} 1.
\bibitem[Garaud and Garaud 2008]{garaud08}Garaud, P., Garaud, J.-D.: 2008, {\it Mon. Not. Roy. Astron. Soc.} {\bf 391,} 1239.
\bibitem[Gough and McIntyre 1998]{gough98}Gough, D.O., McIntyre, M.E.: 1998, {\it Nature} {\bf 394,} 755.
\bibitem[Guerrero and de Gouveia Dal Pino 2007]{guerrero07}Guerrero, G., de Gouveia Dal Pino, E.\ M.: 2007, {\it Astron. Astrophys.} {\bf 464,} 341.
\bibitem[Jiang, Chatterjee, and Choudhuri 2007]{jcc07}Jiang, J., Chatterjee, P., Choudhuri, A.R.: 2007, {\it Mon. Not. Roy. Astron. Soc.} {\bf 381,} 1527.
\bibitem[Karak 2010]{k10} Karak, B.B.: 2010, {\it Astrophys. J.} {\bf 724,} 1021.
\bibitem[Karak and Choudhuri, 2011]{kc11} Karak, B.B., Choudhuri, A.R.: 2011, {\it Mon. Not. Roy. Astron. Soc.} {\bf 410,} 1503.
\bibitem[Karak and Choudhuri, 2012]{kc12} Karak, B.B., Choudhuri, A.R.: 2012, {\it Solar Phys.}, {\bf 278,} 137.
\bibitem[Karak and Nandy, 2012]{kc12} Karak, B.B., Nandy, D.: 2012, arXiv:1206.2106.
\bibitem[MacGregor and Charbonneau 1999]{slowtach2}MacGregor, K.B., Charbonneau, P.: 1999, {\it Astrophys. J.} {\bf 519,} 911.
\bibitem[Miesch 2007]{miesch07}Miesch, M.S.: 2007, {\it Astrophys. J. Lett.} {\bf 658,} L131.
\bibitem[Nandy and Choudhuri 2002]{nc02} Nandy, D., Choudhuri, A.R.: 2002, {\it Science} {\bf 296,} 1671.
\bibitem[Parfrey and Menou 2007]{parfrey07}Parfrey, K.P., Menou, K.: 2007, {\it Astrophys. J. Lett.} {\bf 667,} L207.
\bibitem[Rudiger and Kitchatinov 1997]{slowtach1}Rudiger, G., Kitchatinov, L.L.: 1997, {\it Astron. Nachr.} {\bf 318,} 273.
\bibitem[Spiegel and Zahn 1992]{spiegel92}Spiegel, E.A., Zahn, J.-P.: 1992, {\it Astron. Astrophys.} {\bf 265,}, 106.
\bibitem[Strugarek, Brun, and Zahn 2011]{strugarek11}Strugarek, A., Brun, A.S., Zahn, J.-P.: 2011, {\it Astron. Astrophys.} {\bf 532,} 34.
\bibitem[Yeates, Nandy, and Mackay 2008]{ynm08}Yeates, A.R., Nandy, D., Mackay, D.H.: 2008, {\it Astrophys. J.} {\bf 673,} 544.
\end{thebibliography}

\end{document}